\documentclass[3p,times]{elsarticle}

\usepackage{ecrc}

\usepackage{epstopdf}

\volume{00}

\firstpage{1}

\journalname{Nuclear Physics A}

\runauth{C. Shen et al.}

\jid{nupha}

\jnltitlelogo{Nuclear Physics A}

\usepackage{graphicx}
\usepackage{amsmath,amssymb}

\usepackage[usenames,dvipsnames,svgnames,table]{xcolor}
\usepackage[normalem]{ulem}

\begin{document}

\begin{frontmatter}



\title{Event-by-event direct photon anisotropic flow in relativistic heavy-ion collisions}

\author[label1]{Chun Shen}
\author[label2]{Jean-Fran\c{c}ois Paquet}
\author[label1]{Jia Liu}
\author[label2]{Gabriel Denicol}
\author[label1]{Ulrich Heinz}
\author[label2]{Charles Gale}

\address[label1]{Department of Physics, The Ohio State University, Columbus, Ohio 43210-1117, USA}
\address[label2]{Department of Physics, McGill University, 3600 University Street, Montreal, Quebec, H3A 2T8, Canada}

\begin{abstract}
We consider directly emitted and hadronic decay photons from event-by-event hydrodynamic simulations. We compute the direct photon anisotropic flow coefficients and compare with recent experimental measurements. We find that it is crucial to include the photon multiplicity as a weighting factor in the definition of $v^\gamma_n$. We also investigate the sensitivity of the direct photon spectrum and elliptic flow to the theoretical uncertainty of the photon emission rate in the quark-hadron transition region and to the pre-equilibrium dynamics of relativistic heavy-ion collisions. 
\end{abstract}


\end{frontmatter}

\section{Introduction}
\label{intro}
Electromagnetic radiation is a clean penetrating probe for relativistic heavy-ion collisions, which can provide information that is not accessible with hadronic observables. Recent measurements show a surprisingly large direct photon elliptic flow, comparable with the elliptic flow observed for charged hadrons, both in Au+Au collisions at RHIC \cite{Adare:2011zr} and Pb+Pb collisions at the LHC \cite{Lohner:2012ct}. These results challenge our current theoretical understanding of thermal photon production \cite{Shen:2013cca,Shen:2013vja}. In this work, we improve several aspects in our event-by-event simulations, all of which give small but measurable contributions to the direct photon spectrum and anisotropic flow and thus must be included in attempts to solve the ``direct photon flow puzzle''.

\section{Examining the definition of direct photon anisotropic flow coefficient in event-by-event simulations}
%
\begin{figure}[ht]
\centering
\begin{tabular}{cc}
  \includegraphics[width=0.42\linewidth]{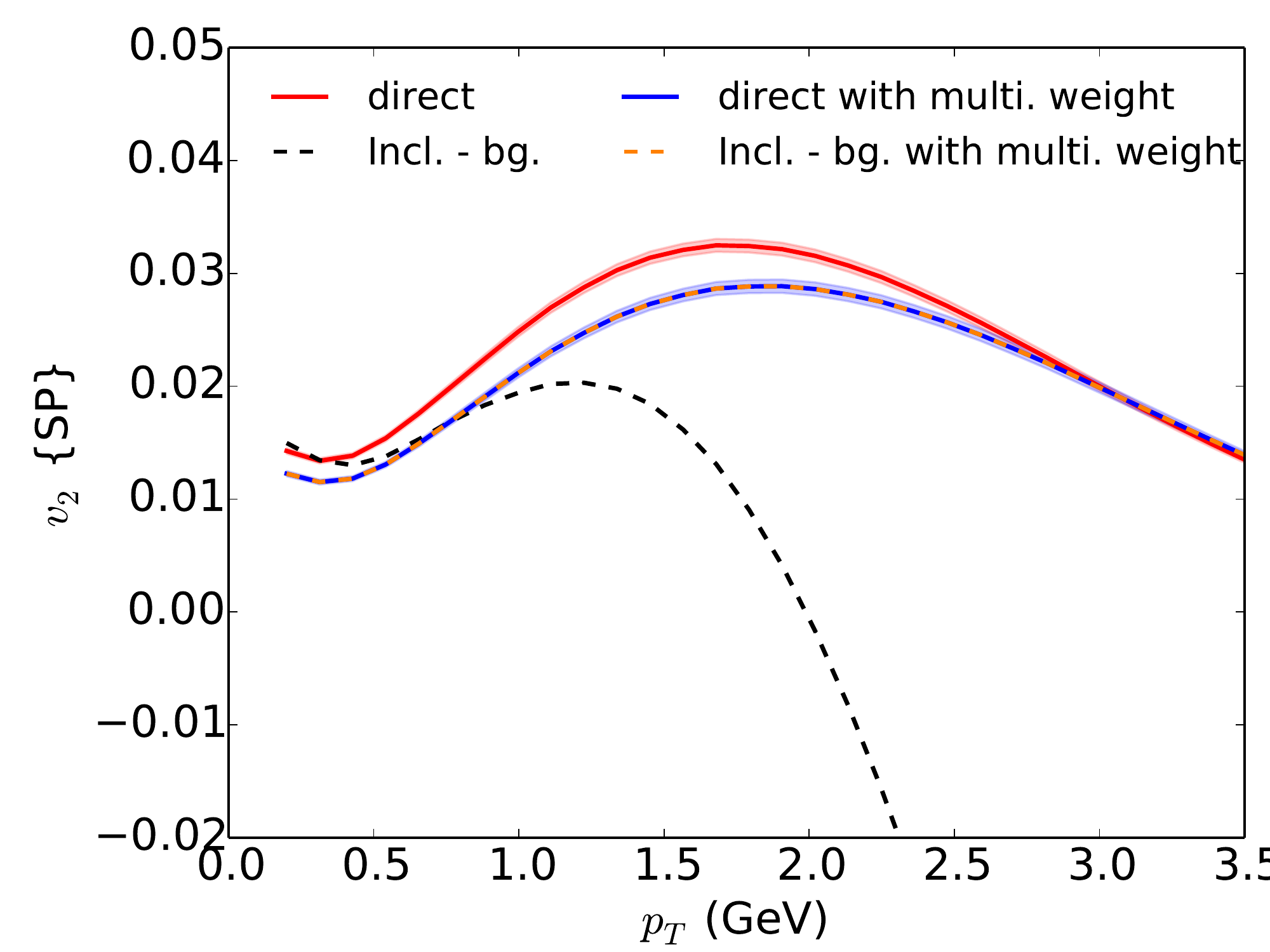} & 
  \includegraphics[width=0.42\linewidth]{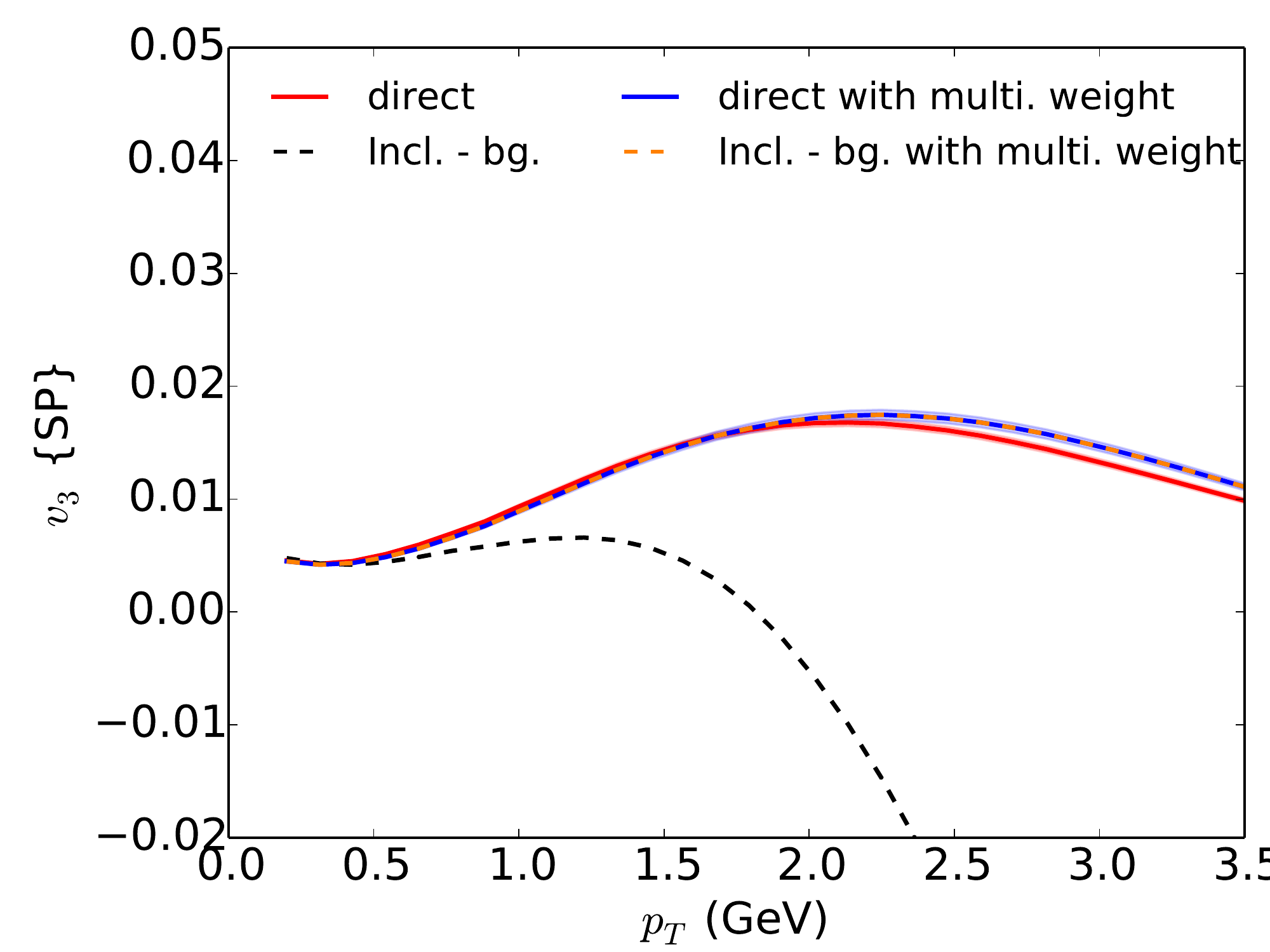}
\vspace*{-3mm}
\end{tabular}
\caption{Comparison between direct photon $v^\gamma_{2,3}\{\mathrm{SP}\}$ and experimentally extracted $v^\gamma_{2,3}\{\mathrm{SP}\}$ from Eqs. (\ref{eq3}) and (\ref{eq4}). Without photon multiplicity weighting the direct and subtraction methods disagree, with the weighting they agree.}
\label{fig1}
\end{figure}
%
In heavy ion collisions, photons are produced through a variety of mechanisms. 
It is not possible to experimentally distinguish all sources of detected photons. Even photons from hadronic decays, which are typically produced on fairly large time scales of the order of $10^7$~fm/$c$, currently cannot be singled out. This is a challenge, since hadronic decay photons completely dominate the photon signal at the RHIC and the LHC. This problem is currently addressed by subtracting from the measured inclusive photons a simulation of the dominant hadronic decays. This procedure effectively defines direct photons in heavy ion collisions.
The anisotropic flow coefficients of direct photons of a single event are given by \cite{Adare:2011zr, Lohner:2012ct}
\begin{equation}
v^{\gamma,\mathrm{dir}}_n (p_T) = \frac{R^\gamma(p_T) v^{\gamma,\mathrm{incl}}_n(p_T) - v^{\gamma,\mathrm{bg}}_n(p_T)}{R^\gamma(p_T) - 1},
\label{eq1}
\end{equation}
where $R^\gamma (p_T) = \frac{dN^{\gamma,\mathrm{incl}}/(dy p_T dp_T)}{ dN^{\gamma,\mathrm{bg}}/(dy p_T dp_T)}$ is the ratio of the measured inclusive photon signal and the simulated decay photon background. Because of the limited number of photons measured in each event, $v^{\gamma, \mathrm{incl}}_n$ and $R^\gamma$ can only be measured as event averages, e.g. $\bar{R}^\gamma= \frac{\langle dN^{\gamma,\mathrm{incl}}/(dy p_T dp_T) \rangle}{\langle dN^{\gamma,\mathrm{bg}}/(dy p_T dp_T)\rangle}$. Whether direct, decay and inclusive photons are still related by Eq. (\ref{eq1}) with event-averaged quantities depends on the precise definition of the photon $v_n$ coefficients. In the following derivations, we will use the scalar-product method $v_n\{\mathrm{SP}\}$ as an example. Similar algebra can be applied to the event plane method. 
If $V_n(p_T) = v_n(p_T) e^{i n \Psi_n(p_T)} = \int d\phi \frac{dN}{dy p_T dp_T d\phi} e^{i n \phi}/(\int d\phi \frac{dN}{dy p_T dp_T d\phi})$ is the underlying complex anisotropic flow vector for one event, the direct photon $v^{\gamma,\mathrm{dir}}_n\{\mathrm{SP}\}$ is defined by
\begin{eqnarray}
  v_{n}^{\gamma , \mathrm{dir}} \{ \mathrm{SP} \} ( p_{T} ) & \equiv & \frac{\mathrm{Re}\{ \langle
  V_{n}^{\gamma , \mathrm{dir}} ( p_{T} ) \cdot ( V_{n}^{\mathrm{ch}} )^*
  \rangle \}}{v_{n}^{\mathrm{ch}} \{ 2 \}} 
  =  \frac{ \mathrm{Re}\left\{ \left\langle \frac{R^{\gamma} ( p_{T} )}{R^{\gamma} ( p_{T} )
  -1} V_{n}^{\gamma , \mathrm{incl}} ( p_{T} ) \cdot ( V_{n}^{\mathrm{ch}}
  )^* \right\rangle - \left\langle \frac{1}{R^{\gamma} ( p_{T} ) -1}
  V_{n}^{\gamma , \mathrm{bg}} ( p_{T} ) \cdot ( V_{n}^{\mathrm{ch}} )^*
  \right\rangle \right\}}{v_{n}^{\mathrm{ch}} \{ 2 \}} \notag \\
   & \neq &  \frac{\bar{R}^{\gamma}( p_{T} )v_{n}^{\gamma ,
  \mathrm{incl}} \{ \mathrm{SP} \} ( p_{T} ) - v_{n}^{\gamma , \mathrm{bg}} \{
  \mathrm{SP} \} ( p_{T} )}{\bar{R}^{\gamma} ( p_{T} ) -1}.
\label{eq3}
\end{eqnarray}
We see that for $v^{\gamma,\mathrm{dir}}_n\{\mathrm{SP}\}$, the direct, decay and inclusive photon anisotropies are not actually related by Eq.~(\ref{eq1}). Now, let us consider the photon multiplicity weighted scalar-product anisotropic flows: 
\begin{eqnarray}
  v_{n}^{\gamma , \mathrm{dir}} \{ \mathrm{SP} , \mathrm{mult} \} ( p_{T} ) & \equiv &
  \frac{\mathrm{Re}\left\{ \left\langle \frac{d N^{\gamma , \mathrm{dir}}}{d y p_{T }
  \mathrm{dp}_{T}} V_{n}^{\gamma , \mathrm{dir}} ( p_{T} ) \cdot (
  V_{n}^{\mathrm{ch}} )^* \right\rangle\right\} }{\left\langle \frac{d N^{\gamma ,
  \mathrm{dir}}}{d y p_{T } \mathrm{dp}_{T}} \right\rangle v_{n}^{\mathrm{ch}} \{ 2
  \}}
   = \frac{ \mathrm{Re}\left\{ \left\langle \left( \frac{d N^{\gamma , \mathrm{incl}}}{d y p_{T
  } \mathrm{dp}_{T}} - \frac{d N^{\gamma , \mathrm{bg}}}{d y p_{T } \mathrm{dp}_{T}}
  \right) \frac{R^{\gamma} ( p_{T} ) V_{n}^{\gamma , \mathrm{incl}} ( p_{T} )
  -V_{n}^{\gamma , \mathrm{bg}} ( p_{T} )}{R^{\gamma} ( p_{T} ) -1} \cdot (
  V_{n}^{\mathrm{ch}} )^*  \right\rangle \right\} }{\left\langle \frac{d N^{\gamma ,
  \mathrm{incl}}}{d y p_{T } \mathrm{dp}_{T}} - \frac{d N^{\gamma , \mathrm{bg}}}{d
  y p_{T } \mathrm{dp}_{T}} \right\rangle v_{n}^{\mathrm{ch}} \{ 2 \}} \notag \\
  & = & \frac{\bar{R}^{\gamma} ( p_{T} ) v_{n}^{\gamma , \mathrm{incl}} \{
  \mathrm{SP} , \mathrm{mult} \} ( p_{T} ) -v_{n}^{\gamma , \mathrm{bg}} \{
  \mathrm{SP} , \mathrm{mult} \} ( p_{T} )}{\bar{R}^{\gamma} ( p_{T} ) -1}.
\label{eq4}
\end{eqnarray}
One sees that with the extra photon multiplicity weight in the definition of the scalar-product flow coefficients, the new quantity $v_n\{\mathrm{SP}, \mathrm{mult}\}$ defined in Eq. (\ref{eq4}) agrees exactly with Eq.~(\ref{eq1}).
In Figure~\ref{fig1}, we numerically verify Eqs.\,(\ref{eq3}) and (\ref{eq4}) for direct photon $v_{2,3}\{\mathrm{SP}\}$ in our event-by-event simulations. Since the photon multiplicity weight factor introduces a bias towards more central collisions which produce more photons per event but less elliptic flow, the final $v^{\gamma, \mathrm{dir}}_2\{\mathrm{SP, mult}\}$ is $\sim$10\% smaller than $v^{\gamma, \mathrm{dir}}_2\{\mathrm{SP}\}$ for a 0-40\% centrality bin.
The result of $v^{\gamma, \mathrm{dir}}_3\{\mathrm{SP, mult}\}$ is close to $v^{\gamma, \mathrm{dir}}_3\{\mathrm{SP}\}$ because in symmetric collisions triangular flow is purely driven by initial state fluctuations and has little centrality dependence. In short, the subtraction procedure shown in Eq.~(\ref{eq1}) only leads to the direct photon $v_n$ if one includes the corresponding multiplicity weights in the event-averaging procedure. 

%
\section{Uncertainty of photon emission rates in the transition region}
In the transition region from QGP to hadron resonance gas (HG), the photon emission rates computed in the QGP phase can be up to a factor of 2 larger than the ones from the HG phase (only photons generated from mesonic interactions are presently considered). In order to avoid a discontinuity in the rates at the transition region, we use the following interpolation:
\begin{equation}
E \frac{d^3 R}{d^3 q} = \frac{T - T_\mathrm{sw, low}}{T_\mathrm{sw,high} - T_\mathrm{sw,low}} \left( E \frac{d^3 R}{d^3 q}\right)_\mathrm{QGP}  + \frac{T_\mathrm{sw, high} - T}{T_\mathrm{sw,high} - T_\mathrm{sw,low}} \left( E \frac{d^3 R}{d^3 q}\right)_\mathrm{HG},
\label{RateTransition.rateeq}
\end{equation}
where $T_\mathrm{sw,low}$ and $T_\mathrm{sw, high}$ are the boundaries of the rate-switching interval.

\begin{figure}[t]
\centering
\begin{tabular}{cc}
  \includegraphics[width=0.42\linewidth]{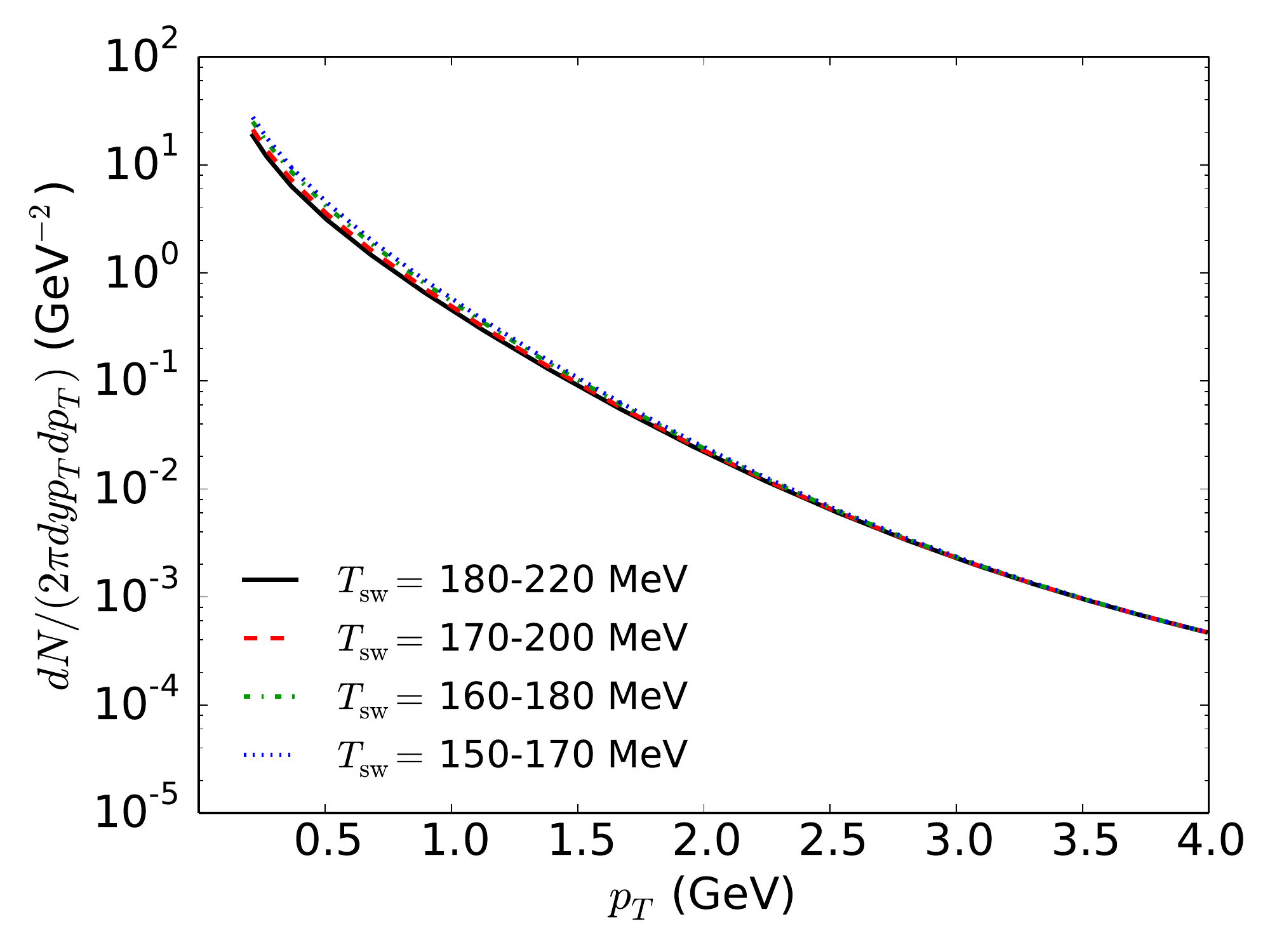} & 
  \includegraphics[width=0.42\linewidth]{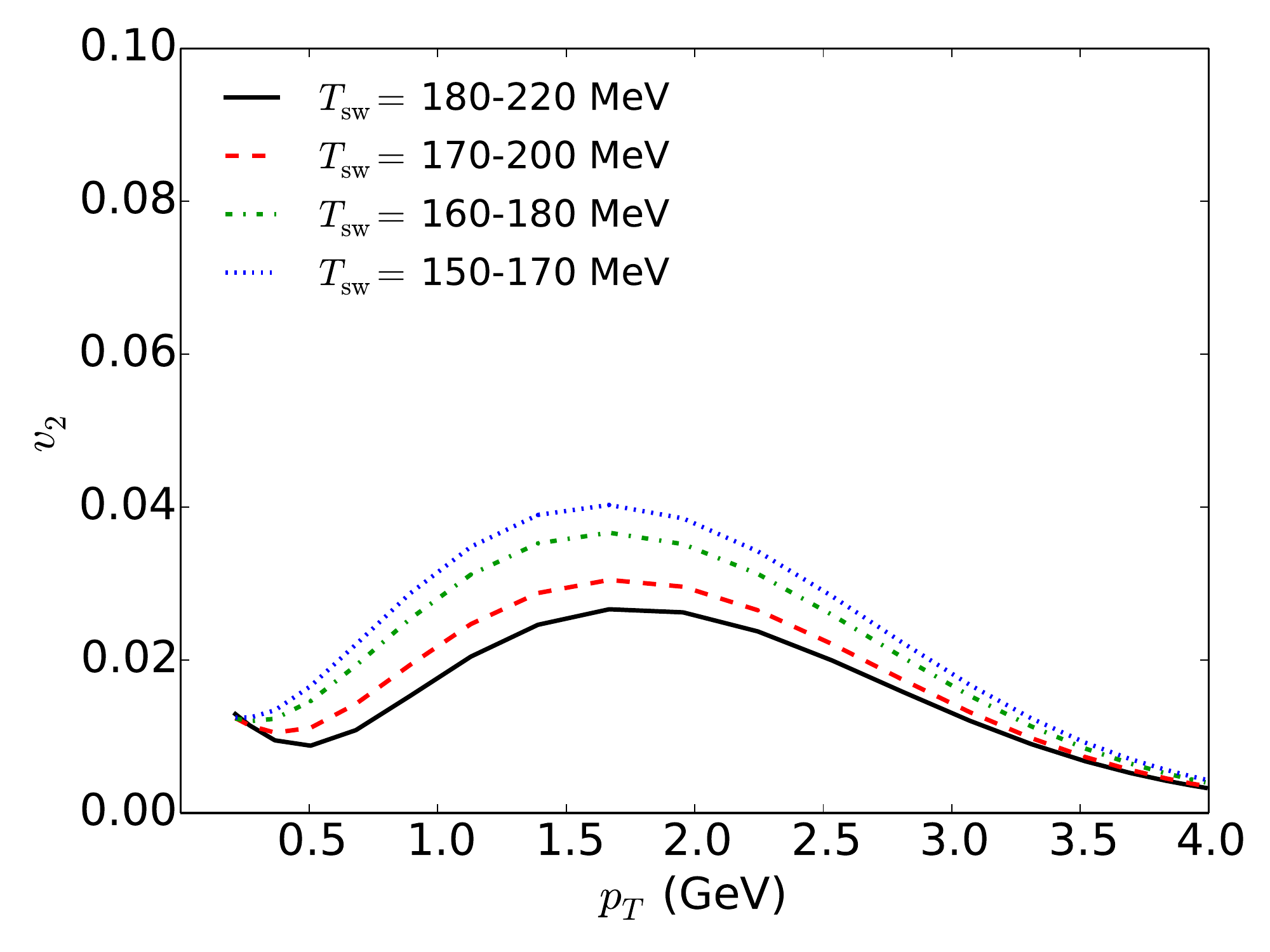}
\vspace*{-2mm}
\end{tabular}
\caption{Direct photon spectra and $v_2$ from hydrodynamic simulations with different transition temperature regions for photon emission rates.}
\label{fig2}
\end{figure}
%
In the left panel of Figure~\ref{fig2}, we study the sensitivity of the direct photon spectra (thermal + pQCD prompt) to different switching temperature regions. With a lower transition temperature, the direct photon spectrum is visibly enhanced at $p_T{\,<\,}2$\,GeV. The $p_T$-integrated thermal photon yield increases by about 30\% for $T_\mathrm{sw}{\,=\,}150{-}170$ MeV compared to switching over the hotter interval $T_\mathrm{sw}{\,=\,}180{-}220$ MeV.
The right panel of Figure~\ref{fig2} shows that the direct photon $v_2$ changes by almost a factor 2 between these two rate switching windows, even after accounting for the dilution of the signal by the prompt pQCD photons which are assumed to carry zero $v_2$. Because the QGP emission rate is higher, more photons are emitted from the transition region when using a lower switching temperature. 
Since the medium anisotropic flow has been almost fully developed in the transition region, the photons emitted from this region carry large anisotropy which contributes significantly to the final direct photon $v_2$. The significant dependence of the direct photon $v_2$ on transition region highlights the uncertainty of the current photon emission rates in the transition region. Nonperturbative techniques \cite{Lin} will be required in order to make progress.

\section{Decay from short lived resonances and pre-equilibrium dynamics}
%
\begin{figure}[h!]
\centering
\begin{tabular}{cc}
  \includegraphics[width=0.42\linewidth]{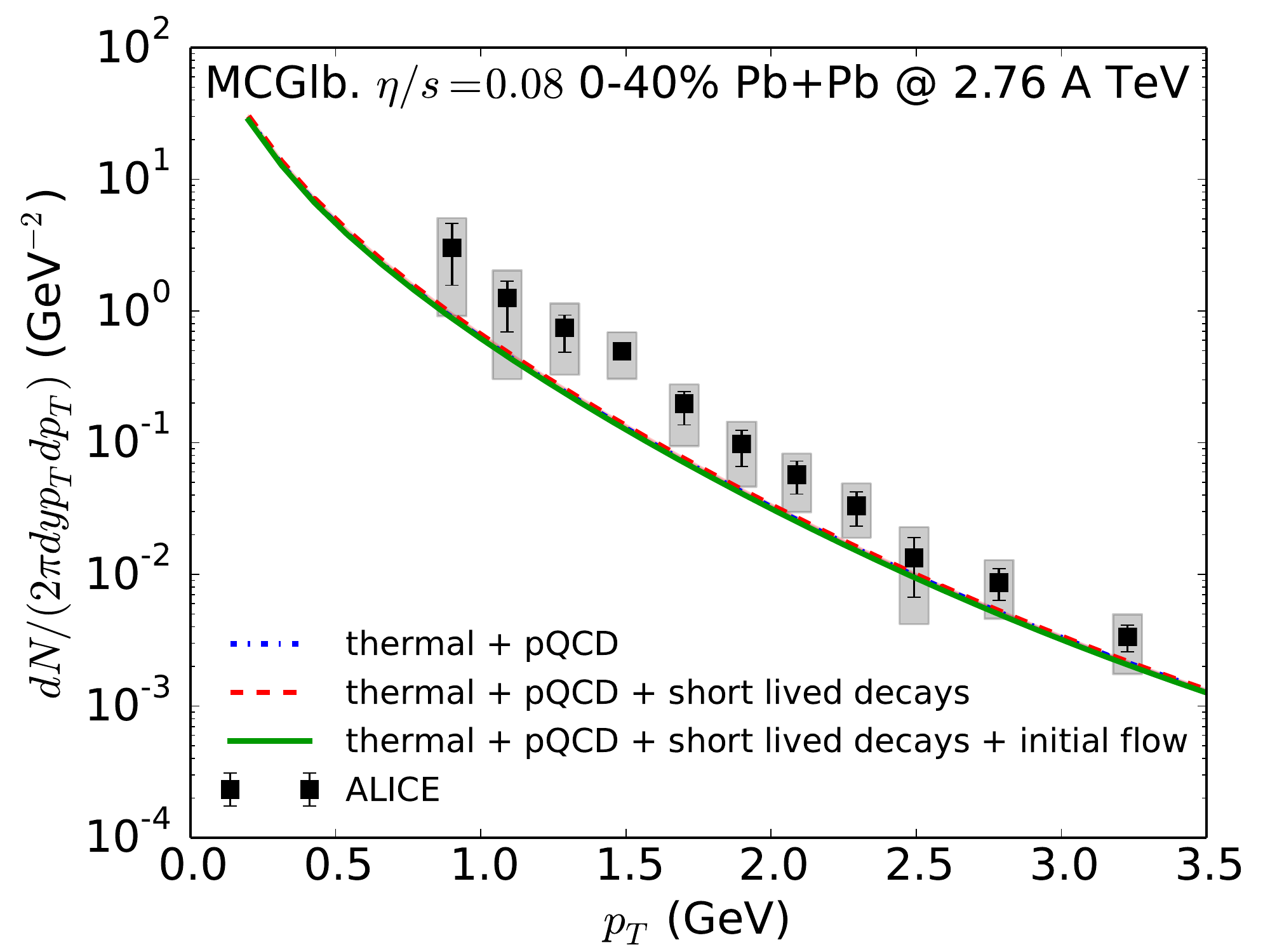} & 
  \includegraphics[width=0.42\linewidth]{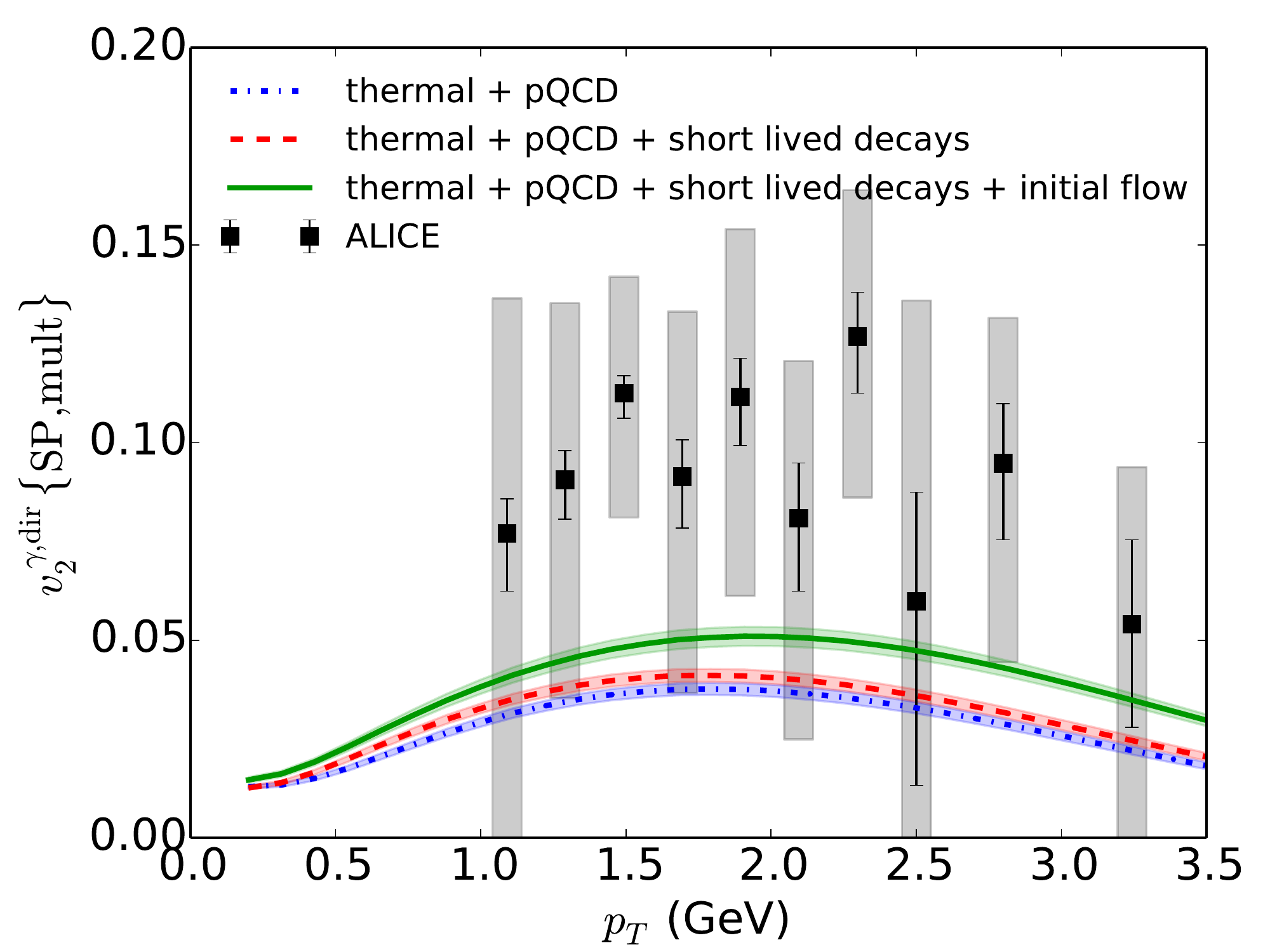} \\
\includegraphics[width=0.42\linewidth]{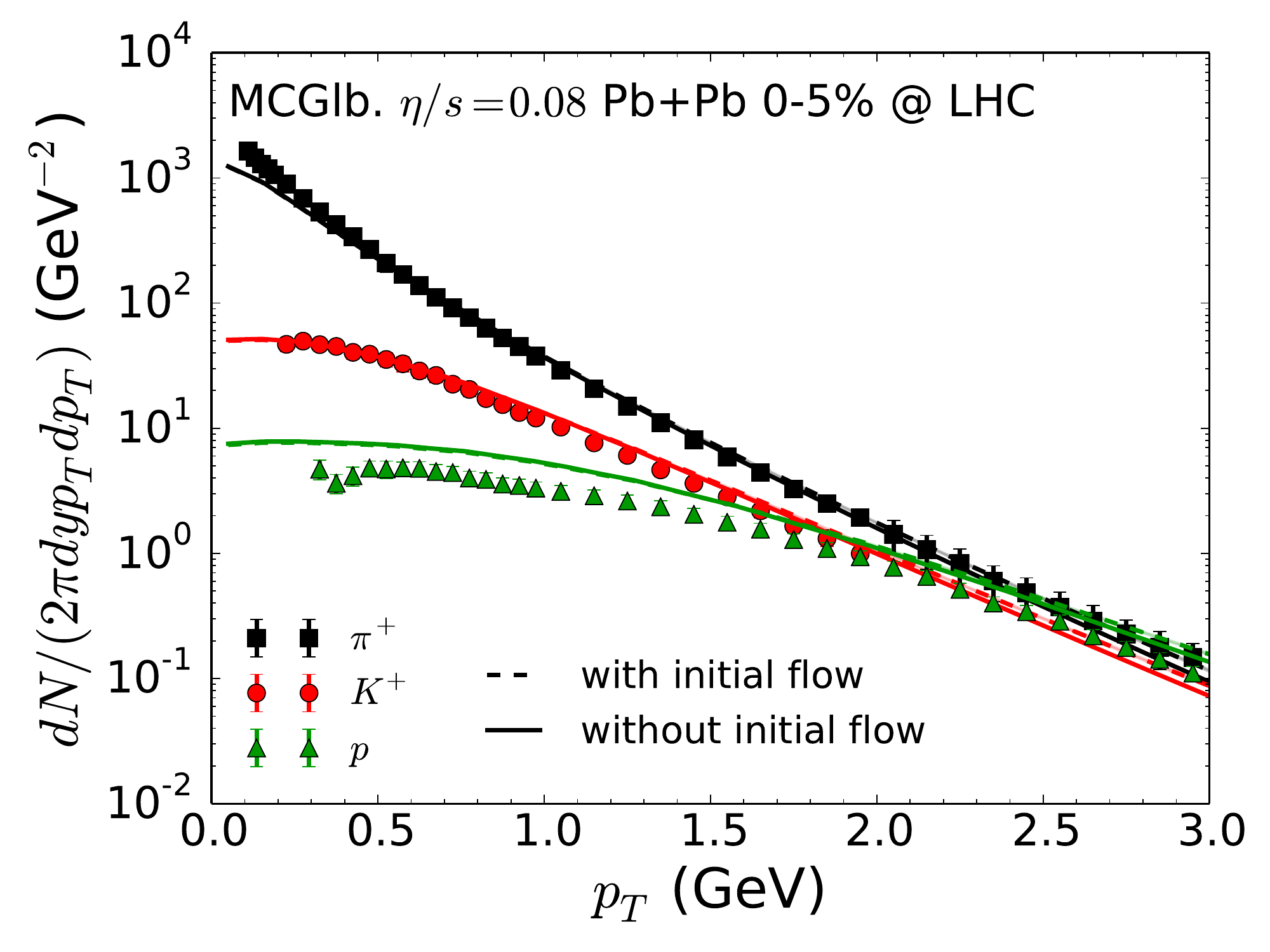} & 
  \includegraphics[width=0.42\linewidth]{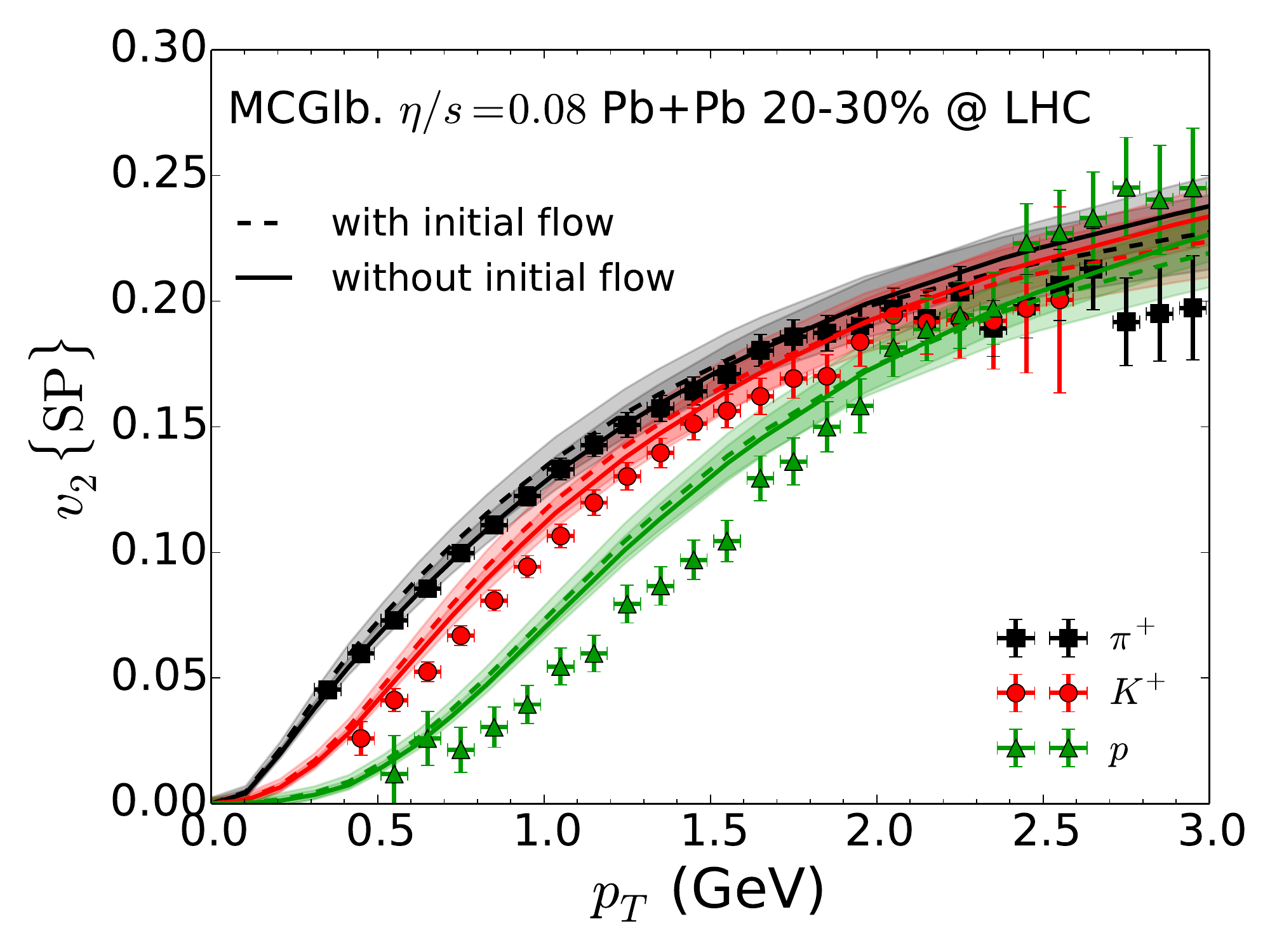}
\end{tabular}
	\caption{{\it Upper panels}: Direct photon spectrum and elliptic flow coefficient compared with ALICE measurements in 2.76\,$A$\,TeV Pb+Pb collisions at 0-40\% centrality \cite{Lohner:2012ct, Wilde:2012wc}. For the transition temperature region for the photon emission rates we take $T_\mathrm{sw}{\,=\,}150{-}170$ MeV. {\it Lower panels}: $\pi^+$, $K^+$, and proton spectra and $v_2$ compared to ALICE data at 0-5\% \cite{Abelev:2013vea} and 20-30\% centrality \cite{Krzewicki:2011ee}, respectively. }
\label{fig3}
\end{figure}
%
Figure~\ref{fig3} shows that our direct photon (thermal + pQCD) spectrum and $v_2\{\mathrm{SP, mult.}\}$ severely underestimate the LHC data \cite{Lohner:2012ct, Wilde:2012wc}.  A similar situation is found for the RHIC data \cite{Adare:2011zr,Shen:2013cca,Bannier:2014ena}. The centrality dependence of direct photon spectra measured at RHIC suggests that we are lacking late hadronic photon emission in the theoretical calculations \cite{Shen:2013vja,Bannier:2014ena}. Prompted by the findings of the recent EMMI Rapid Reaction Task Force on the ``Direct-photon flow puzzle'', we additionally include photons from short-lived resonance decays, which are not included in the experimentally subtracted decay cocktail background. All decay channels for resonances with masses below 2 GeV are included. Although the branching ratio to produce decay photons from each of these unstable resonances is small, many resonances contribute, and overall they increase the direct photon yield by 5-10\% and give a positive albeit small contribution to the direct photon $v_2$. 

Pre-equilibrium dynamics also has the potential to increase the direct photon $v_n$ after thermalization. This is because thermal photon $v_n$ is sensitive to anisotropic hydrodynamic flow during the entire evolution, in contrast to charged hadrons whose momentum distributions only imprint the hydrodynamic flow pattern at kinetic freeze-out. In order to estimate the influence of pre-equilibrium dynamics on the final observables, we use a free-streaming model to evolve the system to $\tau_\mathrm{th}{\,=\,}0.6$\,fm/$c$ where we Landau-match to viscous hydrodynamics. Figure~\ref{fig3} shows that the pre-equilibrium dynamics can significantly increase direct photon elliptic flow because it leads to stronger flow anisotropy during the early hydrodynamic evolution \cite{vanHees:2014ida} without affecting much its final value. In the lower panels of Figure~\ref{fig3} we verify that the pre-equilibrium dynamics assumed here has negligible effects on hadronic flow observables.

\bigskip
\noindent
{\bf Acknowledgments: }
We thank the members of the EMMI RRTF and R. Rapp for fruitful discussions. This work was supported in part by the U.S. Department of Energy under Grants No. \rm{DE-SC0004286} and (within the framework of the JET Collaboration) \rm{DE-SC0004104} and by the Natural Sciences and Engineering Research Council of Canada.


\bibliographystyle{elsarticle-num}
\bibliography{ref.bib}






\end{document}